\documentclass[conference]{IEEEtran}

\IEEEoverridecommandlockouts  

\usepackage[mathscr]{euscript}
{}
{}
\usepackage{nomencl}
\makenomenclature
\usepackage{hyperref}
\usepackage{amsmath}
\usepackage{enumitem}
\usepackage{graphicx,epstopdf}
\usepackage{multicol}
\usepackage{tabularx}
\usepackage{longtable}
\usepackage{textcomp}
\usepackage{multirow}
\usepackage{booktabs}
\usepackage[ruled,vlined]{algorithm2e}
\usepackage{tikz}

\usepackage{color}
\makeatletter
\newcommand*\bigcdot{\mathpalette\bigcdot@{0.5}}
\newcommand*\bigcdot@[2]{\mathbin{\vcenter{\hbox{\scalebox{#2}{$\m@th#1\bullet$}}}}}

\makeatother

\usepackage{amssymb}

\title{Coalitional Game Theory in Power Systems: Applications, Challenges, and Future Directions}

\author{Mukesh Gautam, \emph{Graduate Student Member, IEEE}, Mohammed Benidris, \emph{Senior Member, IEEE}, \\ Department of Electrical \& Biomedical Engineering, University of Nevada, Reno, Reno, NV 89557 \\
(emails: mukesh.gautam@nevada.unr.edu, mbenidris@unr.edu)\vspace{-0.2ex}}

\begin{document}
\maketitle

\begin{abstract}
Game theory-based approaches have recently gained traction in a wide range of applications, importantly in power and energy systems. With the onset of cooperation as a new perspective for solving power system problems, as well as the nature of power system problems, it is now necessary to seek appropriate game theory-based tools that permit the investigation and analysis of the behavior and relationships of various players in power system problems. In this context, this paper performs a literature review on coalitional game theory's most recent advancements and applications in power and energy systems. First, we provide a brief overview of the coalitional game theory's fundamental ideas, current theoretical advancements, and various solution concepts. Second, we examine the recent applications in power and energy systems. Finally, we explore the challenges, limitations, and future research possibilities with applications in power and energy systems in the hopes of furthering the literature by strengthening the applications of coalitional game theory in power and energy systems.
\end{abstract}
\begin{IEEEkeywords}
 Coalitional game theory, energy systems, nucleolus, power systems, Shapley value.
\end{IEEEkeywords}

\section{Introduction}
Game theory-based approaches provide a set of mathematical tools to assess complex interactions and rational behaviors of economic agents in a mutually interactive setting \cite{chen2018resource}. Specifically, coalitional game theory-based approaches have attracted considerable attention of power system researchers because of their ability to uniquely assign payoff among players of the game taking into account their marginal contributions \cite{gautam2021cooperative_freq}.
Coalitional and non-coalitional game theory-based methodologies have been extensively used in a number of power system-related disciplines. Planning, economics, operations, and control of the power system are some of these areas. Game theory-based methods have been applied to determine the optimum dispatch strategies of thermal power plants in \cite{Yildiran15}, coordinate the charging of plug-in hybrid electric vehicles in \cite{7419270}, and allocate the cost of transmission system losses to market participants (customers and power generating companies) in \cite{lima2008cooperative}.

The primary focus of coalitional game theory is the distribution of rewards obtained from player collaboration. When members of a coalition work together and take coordinated action, the collective wealth or value of the coalition is frequently increased or decreased. Naturally, a fascinating and significant issue that has drawn the attention of many mathematicians and scientists is how to distribute the collective reward in an equitable and consistent way. The total payment or incentive of a coalitional game is divided among the players using various solution concepts such the Shapley value, the core, the Nucleolus, and the Nash-bargaining solution.

Applications of coalitional game theoretic approaches in power and energy systems have gained significant momentum in recent years due to their ability to uniquely assign payoffs among players of the game taking into consideration their marginal contributions. A coalitional game theory-based energy management system has been proposed in \cite{querini2020cooperative} to facilitate power exchange of microgrids connected with the main grid.
In \cite{nazari2021economy}, a game theory-based strategy for improving system reliability and reducing power loss in active distribution networks and microgrids has been proposed, where the locational marginal cost was computed at each bus, and each player in the game received financial rewards when system reliability level was improved and the network power losses were reduced. In \cite{gautam2022coop_SRF}, a coalitional game theory-based strategy has been presented for involvement of distributed energy resources in the distribution network to take part in secondary frequency regulation. To determine the optimal locations and sizes of distributed energy resources in distribution systems, a methodology based on coalitional game theory has been presented in \cite{gautam2022coop_DER}. In \cite{bidgoli2022multi}, a multi-stage optimum planning of multi-microgrids utilizing deep learning and coalitional game theory has been developed, where a deep neural network was  utilized for forecasting, and coalitional game theory was used to determine optimal set points for the multi-microgrid. 
A two-layer game model was developed in \cite{yang2021two} to enhance the integrated energy system, where an upper-level Stackelberg game model of the improved energy network was firstly optimized and a cooperative game model for the users, the supply system, and the integrated energy system was developed to carry out an internal optimization. The viability of peer-to-peer (P2P) energy trading in a grid-connected system with voltage limits has been investigated in \cite{azim2021coalition}, where a local voltage control strategy that considers network constraints and prompts prosumers to engage in energy trading has been proposed. Prosumers participating in the P2P energy trading proposed in \cite{azim2021coalition} could join a coalition to discuss and determine the energy trading specifications, including trading volumes and pricing, under a coalitional game-based structure.

Authors of \cite{churkin2021review} have conducted a survey of coalitional game theory applications focusing on expansion planning of power systems. Contrary to the aforementioned paper, this paper presents a review of recent advancements and applications of coalitional game theory in overall power and energy systems. This paper starts by giving the basic introduction of coalitional games along with various solution concepts including the core, Shapley value, nucleolus, and Nash bargaining solutions. Then, the recent advancements of coalitional game theory in both transmission and distribution systems are explained. Moreover, the paper examines the applicability, challenges, and limitations of coalitional game theory through a case study in a 33-node distribution system considering the reserve allocation problem in active distribution systems.

The remainder of the paper is organized as follows: Section~\ref{sec:game} present an introduction to coalitional game theory along with various solution concepts; Section~\ref{sec:app} presents various applications of coalitional game theory in power and energy systems; Section~\ref{sec:case} presents a case study to explain applicability, challenges, and limitations of coalitional game theory; and Section~\ref{sec:conc} summarizes the paper, and presents conclusion and future research directions.

\section{Coalitional Game Theory}\label{sec:game}
In game theory, games are generally categorized into two groups: (a) coalitional games and (b) non-coalitional games. In non-coalitional games, there is no coalition or cooperation between players and they compete among each other to optimize their individual utility functions, while in coalitional games the players form alliances or coalitions with each other to optimize both individual and coalitional utility functions. A coalition must always yield utilities that are equal to or greater than the individual player's utilities since members establish coalitions to optimize their individual utility functions. Non-coalitional games focus mainly on maximizing individual utilities of the players, while coalitional games focus on improving joint utility of the coalition \cite{tan2021fair}. A coalitional game is described by providing a value to each coalition. The following two elements comprise the coalitional game:
\begin{enumerate}
    \item A set of players $\mathscr{N}$, also referred to as the grand coalition.
    \item A characteristic function $V(S):2^{\mathscr{N}} \rightarrow \mathbb{R}$ that converts the set of all feasible player coalitions into a set of coalitional worths or values satisfying the condition $V(\phi)=0$.
\end{enumerate}

Every coalitional game specifies the characteristic function that represents the worths or values of all coalitions. The total value of all members of a coalition serves as the characteristic function of the coalition. Solution paradigms such as the core, the Nucleolus, and the Shapley value are the most popular ways used to allocate the overall payout or incentive among individual players of a coalitional game. 

\subsection{Core of a Coalitional Game}
 In game theory, the core is the set of possible assignments that cannot be enhanced more through any alternative coalitions. The core is a set of payout assignments that ensures no player or player group has a motivation to quit $\mathscr{N}$ to establish a new coalition. Mathematically, the core is defined as follows \cite{shapley1975competitive}. 
 \begin{equation}
     \mathcal{C} = \left\{ \alpha: \sum_{j\in \mathscr{N}} \alpha_j =V(\mathscr{N}) \text{ and } \sum_{j\in S} \alpha_j \geq V(S), \forall S \subset \mathscr{N} \right\}
 \end{equation}

There is no certainty that the cores of coalitional games will always exist. In many cases, the core is in fact empty, making it impossible to stabilize the grand coalition \cite{saad2009coalitional}. Moreover, the core doesn't always give a unique solution and in many cases, the payoff distribution based on the core can be unfair to some players \cite{saad2009coalitional}.  Shapley value or some alternate solution concept may be applied in these circumstances.

\subsection{Nucleolus}
The nucleolus is another important concept in coalitional game theory introduced by Schmeidler \cite{schmeidler1969nucleolus} in 1969. It is founded on the idea of reducing the dissatisfaction of coalition(s) starting with the most dissatisfied coalition(s) \cite{schmeidler1969nucleolus}. The excess of a coalition is the difference between the sum of actual payoffs received by players in the coalition and the worth or value of the coalition. Nucleolus is defined as a payoff distribution vector $\mathbf{x}$ such that the excess (given by \eqref{eqn:excess}) of any potential coalition cannot be lowered without raising any other higher excess.
\begin{equation}
    e_S(\mathbf{x})=V(S)-\sum_{j \in S} x_j \mbox{,} \label{eqn:excess}
\end{equation}
where $\sum_{j \in S} x_j$ denotes the actual value of total payoff received by the players of coalition $S$ and $V(S)$ denotes the worth or value of coalition $S$.

If the core is not empty, the nucleolus should lie in the core as well, guaranteeing the grand coalition's stability \cite{schmeidler1969nucleolus}. However, obtaining the nucleolus might not be simple because of numerical issues \cite{luo2022core}. Additionally, none of the monotonicity requirements are guaranteed \cite{luo2022core}. Moreover, the payoff distribution based on the concept of the nucleolus can be unstable if the core doesn't exist. Therefore, when adopting the nucleolus, it might still be essential to ensure that the core isn't empty.

\subsection{Shapley Value}
The Shapley value is an approach to getting solutions of coalitional game theory. In other words, the Shapley value is a method of distributing the total payoff to each player when everyone plays the game.
The Shapley value is mathematically represented as follows \cite{curiel2013cooperative}.
\begin{equation}
    \psi_j(V) =\hspace{-1.5ex} \sum_{S \in 2^{\mathscr{N}}, j \in S}\hspace{-1.5ex} \frac{(\lvert S\rvert-1)!(n-\lvert S\rvert)!}{n!}[V(S)-V(S\backslash\{j\})]
    \label{eqn:shapley}
\end{equation}
where $n=\lvert \mathscr{N} \rvert$ denotes the total number of players; $S$ is a coalition that is a subset of $\mathcal{N}$; $S\backslash\{j\})$ is a coalition set that excludes player $j$; and $2^{\mathcal{N}}$ is a set of possible coalitions. 

The Shapley value has a number of important properties, which are listed below:
\begin{enumerate}
    \item \textit{Efficiency:} The grand coalition's value is equal to the total of all players' Shapley values, therefore all gains are allocated among the players. Mathematically,
    \begin{equation}
        \sum_{j \in \mathscr{N}} \psi_j(V) = V(\mathscr{N})
    \end{equation}
    \item \textit{Individual Rationality:} When a player participates in a coalition, then its Shapley value should exceed its individual value. Mathematically,
    \begin{equation}
       \psi_j(V) \geq V(\{j\}), \forall j \in \mathscr{N}
    \end{equation}
    \item \textit{Symmetricity:} When two players in a coalition make the same contribution, their Shapley values must be equal. Mathematically, for two players $i$ and $j$ satisfying $V(S \cup \{i\}) = V(S \cup \{j\})$ for each coalition $S$ without $i$ and $j$,
    \begin{equation}
       \psi_i(V) = \psi_j(V) 
    \end{equation}
    \item \textit{Dumminess:} When a player does not increase the coalition's value, its Shapley value should be zero. Mathematically, for player $i$ satisfying $V(S) = V(S \cup \{i\})$ for each coalition $S$ without $i$,
    \begin{equation}
       \psi_i(V) = 0 
    \end{equation}
    \item \textit{Linearity:} The Shapley value corresponding to the sum of characteristic functions equals the sum of Shapley values corresponding to the individual characteristic functions. Mathematically, for two characteristic functions $V_1$ and $V_2$ of a coalitional game,
    \begin{equation}
        \psi(V_1+V_2) = \psi(V_1) + \psi(V_2)
    \end{equation}
\end{enumerate}

\section{Applications of Coalitional Game Theory in Power and Energy Systems}\label{sec:app}
This section presents some of the major applications of coalitional game theory in power and energy systems including transmission and distribution systems operations and planning. 
\subsection{Loss Reduction Allocation of Distributed Generators}
A cooperative game theory-based approach has been implemented in \cite{shaloudegi2012novel} for loss reduction allocation of distributed generators using the 
Shapley values (a solution concept in cooperative game theory). The paper has presented a new locational marginal pricing (LMP) strategy for distribution systems with substantial integration of distributed generation in a competitive energy market. The objective of the LMP technique presented in the paper was to reward distributed generators for their contribution to lower power loss in distribution systems resulting from the participation of all distributed generators in providing demand. Additionally, an iteration-based algorithm has been integrated with the proposed approach. In contrast to the existing LMP-based methodologies, the method presented in the paper has been supposed to offer distribution companies an effective tool for estimating the system state.

\subsection{Reliability-centered Maintenance}
The concept of Shapley value has been utilized in \cite{pourahmadi2016identification} to determine the critical components of the system for reliability-centered maintenance. The paper has stated that the prevention of potential disruptions and the provision of the necessary electricity for end-use customers of distribution networks were two major functions of the generation side of power systems. Since generator failures may have severe or minimal effects depending on the location of generators and network structure, the paper has considered both of these factors during reliability-centered maintenance. In addition to attempting a fair distribution of outage implications to the participating units in the case of $N-k$ contingencies, this study has been primarily focused with the proposal of an index that could be used to sort out the generators based on their outage impacts on power system reliability. The proposed approach has employed Shapley Value to rank generators based on how their failure will affect the reliability of the system. The proposed approach can be used to identify the critical generators of the system, after which the reliability-centered maintenance can be successfully carried out.

\subsection{Under-frequency Load Shedding}
A coalitional game theory-based approach has been proposed in \cite{gautam2021cooperative} for under frequency load shedding control, where a real-time digital simulator has been utilized to compute rate of change of frequency (RoCoF). In order to efficiently and accurately calculate the locations and amounts of loads that need to be shed in order to regulate under frequency load shedding, the paper has provided a two-stage strategy based on cooperative game theory. Using the initial RoCoF referred to the equivalent inertial center, the total amount of loads to be shed, also known as the deficit in generation or the disturbance power, was calculated in the first step. In the second step, load shedding amounts and locations were determined using the Shapley value. The Western Electricity Coordinating Council (WECC) 9-bus 3-machine system has been used to implement the proposed approach, and real-time digital simulators were used for simulation. The findings demonstrated that the proposed under-frequency load shedding technique can successfully restore the system to its pre-disturbance state.

\subsection{Formulation of Distributed Slack Buses}
A coalitional game theory-based method for calculating the participation factors of distributed slack bus generators has been proposed in \cite{gautam2022segan}. The effectiveness of the proposed method has been shown by comparing it to a traditional method for computing the participation factors of distributed slack bus generators. The paper's proposed methodology was a two-stage strategy. The worth (or value) of each participant generator and the coalitions they were a member of was calculated in the first phase. The Shapley value was employed in the second phase to establish participation criteria for each participating generator. The mismatch power was then divided among the several generators using the participation factors. Case studies on the IEEE 14-bus, IEEE 30-bus, and IEEE 57-bus systems were used to show the feasibility of the proposed methodology. The findings demonstrated that systems with distributed slack buses, as opposed to those with a single slack bus, have lower generation costs and power losses.

\subsection{Sizing and Siting of Distributed Energy Resources}
For the purpose of sizing and siting distributed energy resources (DERs), a coalitional game-theoretic strategy has been developed in \cite{gautam2021cooperative_der}. The k-means method has been utilized to perform scenario reduction before identifying potential locations for the deployment of DERs. The two-stage method for choosing the best DER sites and sizes was presented in the paper. Using the equivalent locational marginal prices (LMPs) per unit active power at each bus, a certain number of prospective locations for DERs were chosen in the first stage, and the worths of each prospective location and their coalitions were calculated. The weighted average of LMPs for reduced sets of load scenarios was used to calculate the equivalent LMPs. The Shapley value was employed in the second stage to obtain the optimum placements and sizes for DERs. Case studies performed on several IEEE systems demonstrated that employing the proposed approach as opposed to the existing approaches lowered the total cost of generation after DER deployment.

\subsection{Integrated DER Energy Management}
An integrated DER energy management strategy built on nucleolus estimate has been proposed in \cite{han2020scaling}. The paper has presented a strategy to quickly estimate the nucleolus by including the k-means clustering method, which uses a distinctive marginal allocation pattern as the clustering features. The nucleolus, a method of distributing these economic rewards, has been shown to guarantee the DERs' motivation to participate in coalitional games. The increase in computational time of nucleolus with increase in number of players imposes a hard limit on the system's scalability. A proportional random sampling approach has been proposed for performance assessment. The estimation performances of different clustering algorithms were compared after pairing with different clustering characteristics.

\subsection{Transmission Expansion Planning}
A mechanism for the distribution of transmission expansion expenses among energy market players based on the core and nucleolus paradigms of coalitional games has been proposed in \cite{erli2005transmission}.  Using a coalitional game for a certain number of participants, the cost of transmission line expansion was divided among the participants. Transmission line expansion has been expected to ease transmission congestion, which is believed to be the transmission impediment. The proposed cost allocation was demonstrated through an illustrative case study consisting of an energy market model.

\subsection{Small-signal Stability of Power Systems}
In order to determine which factors are most important for assessing a small-signal stability of a power system, the paper \cite{hasan2018application} has examined several coalitional game theoretic models.
Identifying the factors that have the greatest impact on system stability will make it easier to operate and manage a power system economically in general since network operators and relevant parties would have to put less effort into network management, regulation, and modeling.
The scarce resources might be properly allocated and prioritized after determining which factors have the most influence on small-signal stability concerns.
In contrast to prior methods, a priority ranking algorithm based on a coalitional game theoretic approach has the benefit of taking into account both the individual and all potential cumulative impacts of players. 
In this paper \cite{hasan2018application},  the most significant players have been found using a multi-level strategy that accounted for the network power flow, small-signal stability characteristics, and individual and coalitional behaviors of players.

\subsection{Pre-positioning of Movable Energy Resources}
To enhance the resilience of the power supply, a strategy based on the combination of graph theory and coalitional game theory to determine pre-positioning locations of movable energy resources (MERs) has been presented  in \cite{gautam2022pre}.
The proposed method has been used to identify MERs' pre-positioning sites using weather forecast information to guarantee the quickest response feasible in the case of a natural disaster.
Numerous line outage scenarios were created in the paper using the distribution lines' fragility curves, and a scenario reduction approach was then used to generate a collection of reduced line outage scenarios.
For each reduced line outage scenario, the power distribution network was reconfigured using a graph theory-based approach. The amount of curtailed critical loads and the probability of each reduced line outage scenario have been used to determine the expected load curtailment (ELC) associated to each site.
The Dijkstra shortest path method was used to calculate the best path to take in order to travel to each site.
Using the ELC and the optimum route, MER dispatch cost was calculated.
The potential sites for MER pre-positioning were using the MER dispatch costs.
The size of MER at each prospective location has been determined using the Shapley value.
A 33-node distribution system test bed has been used to verify the proposed method for pre-positioning of MERs. 

\subsection{Microgrid Cooperative Energy Management}
For the cooperative energy trading management of microgrids, a hybrid Energy Management System (EMS) framework based on coalitional games has been developed in \cite{querini2020cooperative}. The paper has determined the energy trading plan that aims to minimize network power losses by using an efficient and robust non-linear model and a characteristic function that ensures the grand coalition is the right coalition structure. A collaborative plan for a moving horizon has been created using the scheduling procedure. The monitoring procedure has employed performance thresholds to track program implementation and detect disturbances in order to improve the effectiveness of microgrids. To protect the privacy of microgrids, each regional EMS was supposed to shares its state summaries with the centralized EMS, which describe its power surplus or deficit for each time frame of the planning horizon. The incentive allocation mechanism distributed the power loss that was acquired as a result of the energy traded during the implementation of the plan using a core-based paradigm. After comparing the proposed framework with a coalition forming game-based methodology reported in the literature, it has been deduced that the task of managing the power exchange between microgrids must be represented as a canonical coalitional game rather than a coalition forming game.

\subsection{Peer-to-Peer Energy Trading}
A framework based on coalitional game theory has been proposed in \cite{malik2022priority} to speed up the development of reliable trading strategies and to motivate participants. Depending on factors like region, peak energy consumption, peak energy generation, and price mechanism, the proposed trading methodology has been designed to provide different preferences at each timespan. The paper has formulated a grand coalition with a goal to enhance the total social welfare and make sure that all players of the game benefit from the policy. Due to the fact that neither peer wishes to initiate a merge or split with respect to its current state, the grand coalition formed by the coalitional game in the paper satisfied Nash equilibrium criteria. The results in \cite{malik2022priority} demonstrated that utilizing the optimal preference for each timespan was preferable than using a single preference for the whole day. The paper has also performed  an economic analysis to determine how to fairly allocate the total payoff among all players of the game. When applying the proposed strategy, customers save money on their energy bills and prosumers earn high profit, according to the economic analysis performed in the paper.

\section{Reserve Allocation in Active Distribution Systems: a Case Study}\label{sec:case}
\begin{figure}
    \centering
    \includegraphics[scale=0.75]{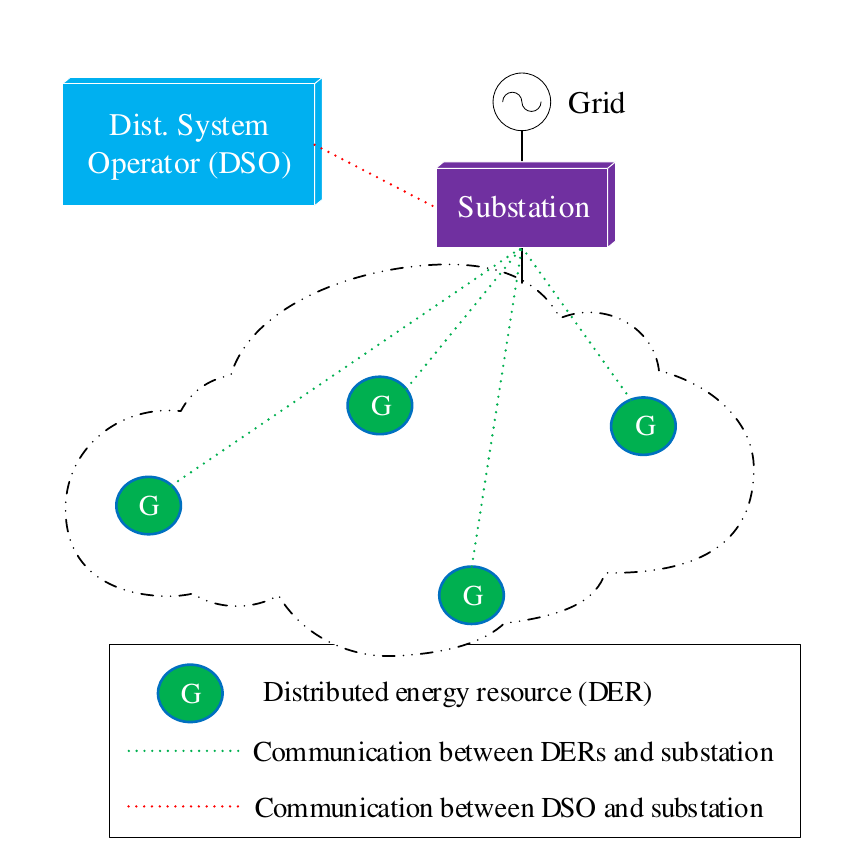}
    \vspace{-1.5ex}
    \caption{Layout of the Coalitional Game Theoretic Model for Reserve Allocation \cite{gautam2022allocating}}
    \vspace{-1.0ex}
    \label{fig:coop_game}
\end{figure}

A case study on reserve allocation in active distribution system based on \cite{gautam2022allocating} is presented to explain applicability, challenges, and limitations of coalitional game theory in power and energy systems. The layout of the coalitional game theory-based approach proposed in \cite{gautam2022allocating} is as shown in Fig.~\ref{fig:coop_game}. A 33-node distribution system with total active and reactive power loads of, respectively, 3715 kW and 2300 kVAr, is considered for the case study. 

\begin{figure}
    \centering  \includegraphics[scale=0.60]{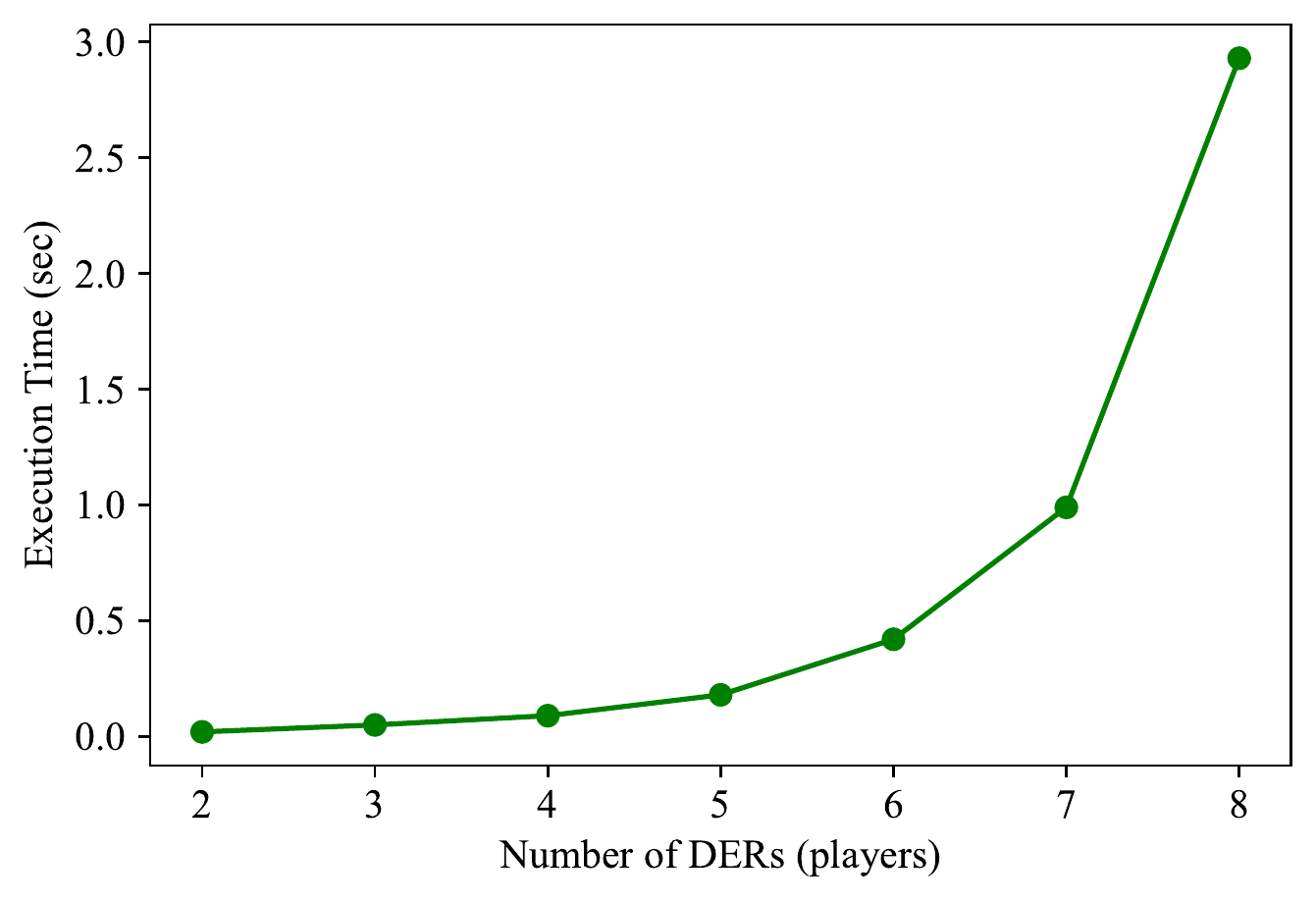}
    \vspace{-2.5ex}\caption{Plot of execution time versus number of DERs} 
    \vspace{-2.5ex}\label{fig:exe_time}
\end{figure}

To analyze the scalability of coalitional game theory-based approaches, two different types of characteristic functions (worthiness index and power loss reduction) are computed by varying the number of DERs (players) from 2 to 8. The equivalent Shapley value is computed for each scenario and distribution factors are then determined for reserve allocation among DERs. Table~\ref{tab:exe_time} shows the number of DERs, their locations, and execution time when the study is conducted on a 64-bit Intel i5 personal computer with processor speed of 3.15 GHz, 8 GB RAM, and Windows operating system. Similarly, Fig.~\ref{fig:exe_time} shows the plot of execution time as the number of DERs is increased. The figure shows that the execution time increases exponentially as the number of DERs (i.e., players of the coalitional game) increases, demonstrating that the coalitional game theory-based approaches are highly unscalable. The coalitional game theory-based approaches are, therefore, highly applicable in power system planning problems where execution time is not a major concern compared to power system operations and control problems. 

\begingroup
\setlength{\tabcolsep}{6pt} 
\renewcommand{\arraystretch}{1.15} 
\begin{table}[]
\centering
\caption{Comparison of Execution Time With Increase in DERs}
\vspace{-1.5ex}
\begin{tabular}{|c|c|c|}
\hline
\begin{tabular}[c]{@{}c@{}}Number of \\ DERs (players)\end{tabular} & \begin{tabular}[c]{@{}c@{}}DER locations\\ (nodes)\end{tabular} & \begin{tabular}[c]{@{}c@{}}Execution \\ time (sec)\end{tabular} \\ \hline
2  & 24, 32 & 0.02  \\ \hline
3   & 7, 14, 24  & 0.05  \\ \hline
4   & 7, 14, 24, 32 & 0.09 \\ \hline
5    & 7, 8, 14, 24, 32  & 0.18  \\ \hline
6  & 7, 8, 14, 24, 25, 32  & 0.42  \\ \hline
7   & 7,8,14,24,25,30,32 & 0.99   \\ \hline
8  & 7, 8, 14, 24, 25, 30, 31, 32   & 2.93  \\ \hline
\end{tabular}
\label{tab:exe_time}
\vspace{-2ex}
\end{table}
\endgroup

\section{Conclusion and Future Research Directions}\label{sec:conc}
This paper has presented a review of coalitional game theory applications in power and energy systems. The paper started with the basic introduction to coalitional game theory along with advantages and limitations of various solution concepts of coalitional games including the core, the nucleolus, and the Shapley value. The various applications of coalitional game theory were, then, presented. A case study of reserve allocation in active distribution systems was presented to point out applicability, challenges, and limitations of coalitional game theory in power and energy systems. 

In the case study, the execution time was computed by increasing the number of players. The case study results showed that the execution time increases exponentially as the number of players in coalitional games is increased. It was, therefore, deduced that the coalitional game theory is applicable in case of power system operations and control problems if there are few numbers of players. On the other hand, the coalitional game theory-based applications are still viable in case of power system planning problems where execution time is not a prime concern. The consideration of marginal contribution of each player while distributing the overall reward among all players makes the coalitional game theoretic techniques based on Shapley values still favorable. 

Since the existing approaches for computing the solutions of the coalitional games are not scalable, further research is needed to develop approaches that are computationally efficient and accurate enough. Since most of the solution concepts are based on the enumeration of possible coalition sets and computation of corresponding characteristic functions, the size of the coalition set increases non-linearly with increase in the number of players. Moreover, some coalitions might be more impactful and significant compared to others. Based on this, some approximations can be made while determining the solutions. This can help reduce the computational time while solving a coalitional game.


\bibliographystyle{elsarticle-num}
\bibliography{References.bib}

\begin{thebibliography}{10}
\expandafter\ifx\csname url\endcsname\relax
  \def\url#1{\texttt{#1}}\fi
\expandafter\ifx\csname urlprefix\endcsname\relax\def\urlprefix{URL }\fi
\expandafter\ifx\csname href\endcsname\relax
  \def\href#1#2{#2} \def\path#1{#1}\fi

\bibitem{chen2018resource}
Y.~Chen, B.~Ai, Y.~Niu, K.~Guan, Z.~Han, Resource allocation for
  device-to-device communications underlaying heterogeneous cellular networks
  using coalitional games, IEEE Transactions on Wireless Communications 17~(6)
  (2018) 4163--4176.

\bibitem{gautam2021cooperative_freq}
M.~Gautam, N.~Bhusal, M.~Benidris, H.~Livani, A cooperative game theory-based
  secondary frequency regulation in distribution systems, in: North American
  Power Symposium (NAPS), IEEE, 2021, pp. 1--6.

\bibitem{Yildiran15}
N.~Yildiran, E.~Tacer, Game theory approach to solve economic dispatch problem,
  International Journal of Trade, Economics and Finance 6~(4) (2015) 230--234.

\bibitem{7419270}
J.~{Tan}, L.~{Wang}, A game-theoretic framework for vehicle-to-grid frequency
  regulation considering smart charging mechanism, IEEE Transactions on Smart
  Grid 8~(5) (2017) 2358--2369.

\bibitem{lima2008cooperative}
D.~A. Lima, J.~Contreras, A.~Padilha-Feltrin, A cooperative game theory
  analysis for transmission loss allocation, Electric Power Systems Research
  78~(2) (2008) 264--275.

\bibitem{querini2020cooperative}
P.~L. Querini, O.~Chiotti, E.~Fern{\'a}dez, Cooperative energy management
  system for networked microgrids, Sustainable Energy, Grids and Networks 23
  (2020) 100371.

\bibitem{nazari2021economy}
M.~H. Nazari, M.~B. Sanjareh, A.~Khodadadi, M.~Torkashvand, S.~H. Hosseinian,
  An economy-oriented dg-based scheme for reliability improvement and loss
  reduction of active distribution network based on game-theoretic sharing
  strategy, Sustainable Energy, Grids and Networks 27 (2021) 100514.

\bibitem{gautam2022coop_SRF}
M.~Gautam, N.~Bhusal, M.~Benidris, H.~Livani, A cooperative game theory-based
  secondary frequency regulation in distribution systems, in: 53rd North
  American Power Symposium (NAPS), IEEE, 2022, pp. 1--6.

\bibitem{gautam2022coop_DER}
M.~Gautam, N.~Bhusal, M.~Benidris, A cooperative game theory-based approach to
  sizing and siting of distributed energy resources, in: 2021 53rd North
  American Power Symposium (NAPS), IEEE, 2022, pp. 1--6.

\bibitem{bidgoli2022multi}
M.~A. Bidgoli, A.~Ahmadian, Multi-stage optimal scheduling of multi-microgrids
  using deep-learning artificial neural network and cooperative game approach,
  Energy 239 (2022) 122036.

\bibitem{yang2021two}
S.~Yang, Z.~Tan, J.~Zhou, F.~Xue, H.~Gao, H.~Lin, et~al., A two-level game
  optimal dispatching model for the park integrated energy system considering
  stackelberg and cooperative games, International journal of electrical power
  \& energy systems 130 (2021) 106959.

\bibitem{azim2021coalition}
M.~I. Azim, W.~Tushar, T.~K. Saha, Coalition graph game-based p2p energy
  trading with local voltage management, IEEE Transactions on Smart Grid 12~(5)
  (2021) 4389--4402.

\bibitem{churkin2021review}
A.~Churkin, J.~Bialek, D.~Pozo, E.~Sauma, N.~Korgin, Review of cooperative game
  theory applications in power system expansion planning, Renewable and
  Sustainable Energy Reviews 145 (2021) 111056.

\bibitem{tan2021fair}
M.~Tan, Y.~Zhou, L.~Wang, Y.~Su, B.~Duan, R.~Wang, Fair-efficient energy
  trading for microgrid cluster in an active distribution network, Sustainable
  Energy, Grids and Networks 26 (2021) 100453.

\bibitem{shapley1975competitive}
L.~S. Shapley, M.~Shubik, Competitive outcomes in the cores of market games,
  International Journal of Game Theory 4~(4) (1975) 229--237.

\bibitem{saad2009coalitional}
W.~Saad, Z.~Han, M.~Debbah, A.~Hjorungnes, T.~Basar, Coalitional game theory
  for communication networks, Ieee signal processing magazine 26~(5) (2009)
  77--97.

\bibitem{schmeidler1969nucleolus}
D.~Schmeidler, The nucleolus of a characteristic function game, SIAM Journal on
  applied mathematics 17~(6) (1969) 1163--1170.

\bibitem{luo2022core}
C.~Luo, X.~Zhou, B.~Lev, Core, shapley value, nucleolus and nash bargaining
  solution: A survey of recent developments and applications in operations
  management, Omega (2022) 102638.

\bibitem{curiel2013cooperative}
I.~Curiel, Cooperative game theory and applications: cooperative games arising
  from combinatorial optimization problems, Vol.~16, Springer Science \&
  Business Media, 2013.

\bibitem{shaloudegi2012novel}
K.~Shaloudegi, N.~Madinehi, S.~Hosseinian, H.~A. Abyaneh, A novel policy for
  locational marginal price calculation in distribution systems based on loss
  reduction allocation using game theory, IEEE transactions on power systems
  27~(2) (2012) 811--820.

\bibitem{pourahmadi2016identification}
F.~Pourahmadi, M.~Fotuhi-Firuzabad, P.~Dehghanian, Identification of critical
  components in power systems: A game theory application, in: IEEE Ind. Appl.
  Society Annual Meeting, IEEE, 2016, pp. 1--6.

\bibitem{gautam2021cooperative}
M.~Gautam, N.~Bhusal, M.~Benidris, A cooperative game theory-based approach to
  under-frequency load shedding control, in: IEEE Power \& Energy Society
  General Meeting (PESGM), IEEE, 2021, pp. 1--5.

\bibitem{gautam2022segan}
M.~Gautam, N.~Bhusal, J.~Thapa, M.~Benidris, A cooperative game theory-based
  approach to formulation of distributed slack buses, Sustainable Energy, Grids
  and Networks 32 (2022) 100890.

\bibitem{gautam2021cooperative_der}
M.~Gautam, N.~Bhusal, M.~Benidris, A cooperative game theory-based approach to
  sizing and siting of distributed energy resources, in: 2021 North American
  Power Symposium (NAPS), IEEE, 2021, pp. 01--06.

\bibitem{han2020scaling}
L.~Han, T.~Morstyn, M.~D. McCulloch, Scaling up cooperative game theory-based
  energy management using prosumer clustering, IEEE Transactions on Smart Grid
  12~(1) (2020) 289--300.

\bibitem{erli2005transmission}
G.~Erli, K.~Takahasi, L.~Chen, I.~Kurihara, Transmission expansion cost
  allocation based on cooperative game theory for congestion relief,
  International Journal of Electrical Power \& Energy Systems 27~(1) (2005)
  61--67.

\bibitem{hasan2018application}
K.~N. Hasan, R.~Preece, J.~V. Milanovi{\'c}, Application of game theoretic
  approaches for identification of critical parameters affecting power system
  small-disturbance stability, International Journal of Electrical Power \&
  Energy Systems 97 (2018) 344--352.

\bibitem{gautam2022pre}
M.~Gautam, M.~Benidris, Pre-positioning of movable energy resources for
  distribution system resilience enhancement, in: 2022 International Conference
  on Smart Energy Systems and Technologies (SEST), IEEE, 2022, pp. 1--6.

\bibitem{malik2022priority}
S.~Malik, M.~Duffy, S.~Thakur, B.~Hayes, J.~Breslin, A priority-based approach
  for peer-to-peer energy trading using cooperative game theory in local energy
  community, International Journal of Electrical Power \& Energy Systems 137
  (2022) 107865.

\bibitem{gautam2022allocating}
M.~Gautam, M.~M. Lakouraj, N.~Bhusal, M.~Benidris, H.~Livani, Allocating
  reserves in active distribution systems for tertiary frequency regulation,
  in: 2022 IEEE Power \& Energy Society Innovative Smart Grid Technologies
  Conference (ISGT), IEEE, 2022, pp. 1--5.

\end{thebibliography}

\end{document}